\documentclass[12pt]{article}
\usepackage{graphicx,amsmath,cancel,fullpage,mhchem,lscape}
\usepackage[usenames,dvipsnames]{xcolor}
\usepackage{caption}
\usepackage{subcaption}

\begin{document}

\title{On the Strength of Glasses}
\author{Apiwat Wisitsorasak$^{a,b}$ and Peter G. Wolynes$^{*,a,b,c}$ \\
$^a$ \small Center for Theoretical Biological Physics, $^b$ \small Department of Physics \& Astronomy, \\
$^c$ \small Department of Chemistry, Rice University, Houston, TX \\
\texttt{pwolynes@rice.edu}}
\maketitle
\begin{abstract}
The remarkable strength of glasses is examined using the random first order transition theory of the glass transition. The theory predicts that strength depends on elastic modulus but also on the configurational energy frozen in when the glass is prepared. The stress catalysis of cooperative rearrangements of the type responsible for the supercooled liquid's high viscosity account quantitatively for the measured strength of a range of metallic glasses, silica and a polymer glass.
\end{abstract}

A fundamental question about solid matter is what ultimately determines its mechanical strength. Glasses, in the popular mind, are easy to break but in fact, if surface cracks are carefully avoided, glasses turn out to be intrinsically quite strong. Nearly a century ago, Frenkel provided an elegant argument for the maximum stress that a solid could withstand \cite{frenkel1926theorie}. Crystalline metals were found to be hundreds to thousands of times weaker than the Frenkel estimate \cite{Zhu2010710}. This observation inspired the extremely fruitful ideas of dislocations and grain boundaries that provide easy ways for polycrystalline metals to rearrange and plastically deform \cite{orowan1940problems,taylor1934mechanism,greaves2011poisson,argon1979plastic}. Glasses come much closer to the Frenkel limit but still fall short in strength  \cite{chen2008mechanical}. In this paper we explore quantitatively the notion that the mechanical failure of glassy materials ultimately arises from strain catalyzed rearrangements of the same kind as those responsible for the high supercooled liquid viscosity. The idea that there is a relation of yield strength to the glass transition itself is not new  and has been examined in various ways \cite{nieh2006unified,johnson2005universal,wang2004bulk,argon1979plastic,spaepen1977microscopic,wondraczek2011towards}. Here we go further by exploiting the current quantitative understanding of cooperatively rearranging regions that has emerged from the random first order transition (RFOT) theory of glasses  \cite{kirkpatrick1989scaling,kirkpatrick1987stable,kirkpatrick1987connections,xia2000fragilities,lubchenko2007theory,xia2001microscopic,peter2009spatiotemporal} in order to make some specific predictions. RFOT theory describes the microscopic origin of cooperatively rearranging regions and predicts they are compact containing a few hundred molecular units near the laboratory glass transition temperature $T_g$. These regions become more fractal, resembling strings or percolation clusters \cite{stevenson2006shapes} at higher temperatures where flow is no longer thermally activated \cite{stoessel1984linear} but rather dominantly collisional. The quantitative predictions of RFOT theory concerning the well-established thermodynamic/kinetic correlations in the viscous liquid state, dynamical heterogeneity in supercooled liquids \cite{xia2001microscopic} and the aging \cite{lubchenko2004theory} and rejuvenating \cite{peter2009spatiotemporal} properties of the glassy state proper agree quite well with observations \cite{stevenson2005thermodynamic}. It is thus natural to enquire as to what the theory predicts for the material strength of glasses. 

\begin{figure}[h!] 
  \centering
      \noindent\includegraphics[width=38pc,angle=0]{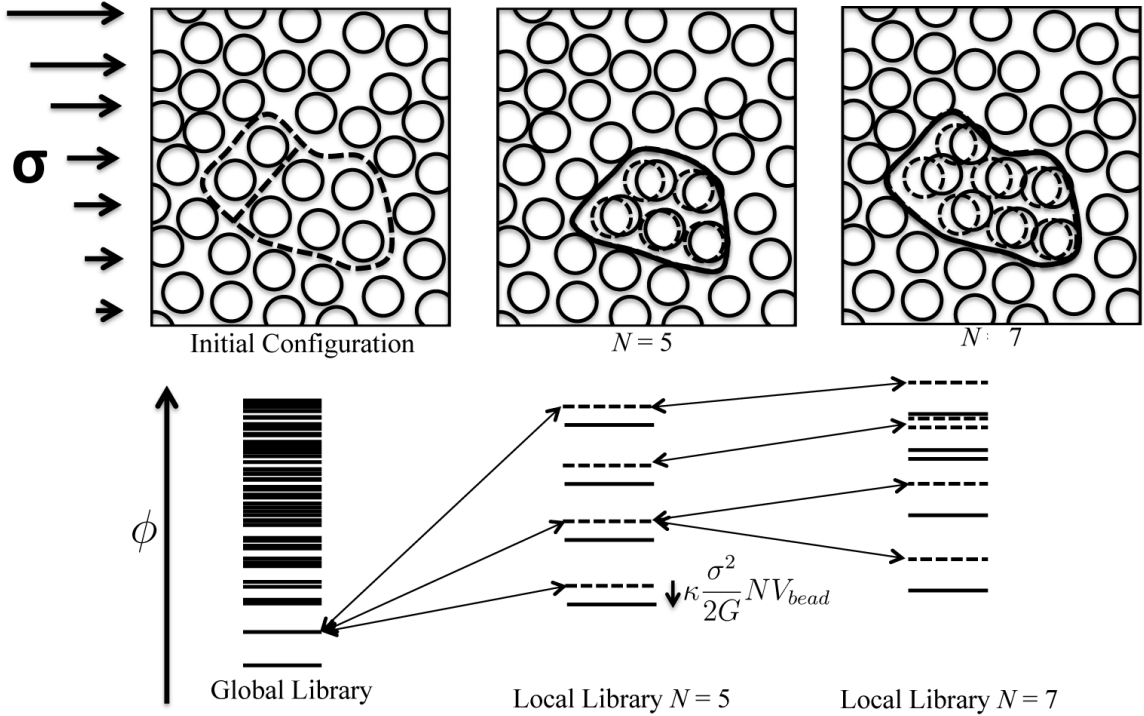} 
  \caption{We show in the upper part of the figure schematic snapshots of local rearrangement starting from an initial frozen configuration in an imposed stress field $\sigma$.  Following Lubchenko and Wolynes \cite{lubchenko2004theory} the lower left panel shows the spectrum of possible free energy minima for a large sample of glass. Levels are  listed in order of the internal free energy $\phi$, comprising the potential energy along with a vibrational entropic contribution. When the glass is trapped in a single such state, local regions of size $N$ can rearrange to new minima while only  weakly disturbing their environment elastically. Connected energy levels are shown in the next two panels. When an imposed stress $\sigma$ is imposed the energy levels are shifted and the energy cost of moving  $N$ particles is reduced by an amount $(\kappa \sigma^2/2 G)N V_{\mathrm{bead}}$ where $G$ is the elastic modulus and $\kappa$  is a factor that includes the elastic response of the environment which does not shift to a new minimum. $V_{\mathrm{bead}}$ is the volume of a molecular unit. Eventually for sufficiently large $N$ a distinct structure is formed coincident in free energy with the initial state, allowing irreversible motions to occur.} \label{fig:LocalLibraryStress}
\end{figure}

\begin{figure}[h!] 
  \centering
      \noindent\includegraphics[width=19pc,angle=0]{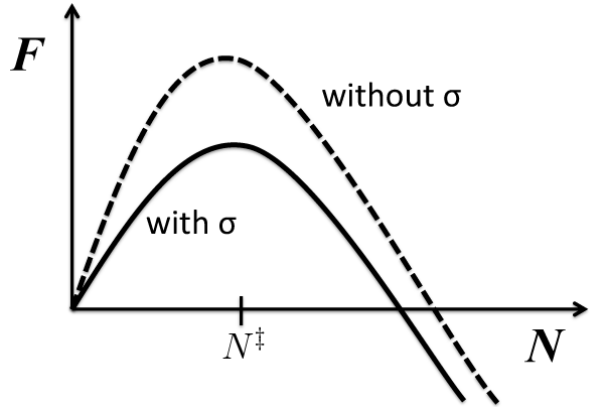} 
  \caption{The local level libraries in Fig.~\ref{fig:LocalLibraryStress} when thermally averaged yield a free energy profile for rearranging a specific region as a function of the number of displaced particles $N$. The average mismatch energy is balanced against a term containing the configurational entropy (from averaging over all the states), any initially excess energy frozen into the glass along with a contribution from relaxing strains via reconfiguration  in an imposed stress field. The activation barrier is lowered by the imposed stress, eventually vanishing when the stress is sufficiently large, leading to rapid failure of the glass's structural integrity.} \label{fig:FreeEnergyPlot}
\end{figure}

We begin by reviewing how activated events occur in liquids and glasses in the absence of stress. The easiest way to conceptualize activated events in the RFOT theory is through what is called the Òlandscape library constructionÓ by Lubchenko and Wolynes  \cite{lubchenko2004theory}. This construction has also been used to define point-to-set correlation lengths \cite{berthier2005direct,bouchaud2004adam} allowing many key points of RFOT theory to be confirmed via computer simulations \cite{cavagna2012dynamic,cavagna2010numerical,cammarota2009evidence,biroli2008thermodynamic}. This construction is schematically pictured in Fig. \ref{fig:LocalLibraryStress}. In mean field RFOT theory, below a dynamical transition temperature $T_A$, the system becomes trapped in one of an exponentially large number of possible metastable states which are minima of a free energy functional \cite{kirkpatrick1987stable}. For molecular fluids these states can be taken as nearly structurally synonymous with the Òinherent structuresÓ which precisely correspond to minima of the potential energy \cite{xia2000fragilities}, but the individual stability of these states at finite temperature depends not only on their energy but also on their vibrational entropy. Irreversible reconfiguration events eventually take place by rearranging molecules in ever larger regions of size $N$ until a critical size is reached. Above the Kauzmann temperature, $T_K$, the configurational entropy is extensive and so as the size of a reconfiguring region increases the number of possible local rearrangements grows as well. Generally moving to any one of these rearranged structures costs free energy because the environment of the rearranging region does not fit the new locally accessible alternative structures as well as it fits the original free energy minimum from which rearrangement starts. The typical mismatch energy $\Delta E(N)$ near the Kauzmann transition scales as $\gamma N^x$. The power law in mean field theory represents a surface energy \cite{kirkpatrick1987stable} so the exponent $x = 2/3$ but scaling arguments \cite{kirkpatrick1989scaling} suggest there should be a somewhat weaker scaling with $x = 1/2$ near an ideal glass transition at $T_K$ due to wetting from the numerous alternative states that can be interpolated between the fixed environment and the core of the rearranging region. 

Xia and Wolynes showed the coefficient in the mismatch energy can be computed near $T_K$ by assuming a locally sharp interface and by making a microscopic estimate using density functional theory of the localization free energy which is entropic:
\begin{eqnarray}
\gamma = \frac{3\sqrt{3 \pi}}{2}k_B T_K \ln \left(\frac{\alpha_L a^2}{\pi e} \right).
\end{eqnarray}
Here $\alpha_L$ determines the size of the vibrational fluctuations in a minimum and is roughly 100, reflecting displacements following Lindemann's stability criterion allowing localized motion of about one tenth of the interparticle spacing, $a$ .

Above $T_K$ any mismatch energy can however be overcome by the entropic driving force favoring reconfiguration to one of the many alternate structures, $F_{\mathrm{bulk}}(N) = - T s_c N$   where $s_c$ is the configurational entropy per particle. Balancing $F_{\mathrm{bulk}}(N)$ and $\Delta E(N)$ gives an activation free energy to be overcome for irreversible rearrangement, $\Delta F^\ddag$, which is a function of $s_c$.  $\Delta F^\ddag$  diverges near $T_K$ as $s_c$ vanishes. This prediction then connects the kinetics of rearrangements with thermodynamics, a hallmark of the RFOT theory. Using the approximate coefficient $\gamma$ obtained by Xia and Wolynes the absolute magnitude of barriers is also predicted to follow an Adam-Gibbs-like relation $\Delta F^\ddag = A/s_c$ but with a specific numerical value for $A = \frac{27 \pi}{16} k_B (\ln (\alpha_L a^2/\pi e))^2$. Because the Lindemann parameter $\alpha_L$ depends only weakly on the potential, in RFOT theory then  $\Delta F^\ddag/k_B T$ is again dominantly a function of the configurational entropy, across a range of substances.

The landscape library argument can also be used in the so-called ``aging" regime to describe motion in the glass \cite{lubchenko2004theory}. In the aging regime, the initial configuration is not one chosen from the thermal equilibrium ensemble at the ambient temperature but instead structurally resembles a system that was equilibrated at a higher so-called ``fictive'' temperature. In a simple quench to low temperature the fictive temperature initially is the laboratory glass transition temperature $T_g$. For this nonequilibrium situation the initial configuration then will not only gain entropy by reconfiguring locally but also will release an additional energy per particle $\Delta \Phi$ which represents the energy frozen in at the glass transition \cite{lubchenko2004theory}.  If we assume the configurational heat capacity has the empirical form $\Delta c_p (T_g) \cdot (T_g/T)$ this excess energy is  $\Delta \Phi = \Delta c_p(T_g) T_g  \ln \frac{T_g}{T} $.

Owing to this excess driving force, reconfiguration events occur sooner in a glass than they would in a liquid structurally equilibrated to the same ambient temperature. The nonequilibrium activation free energy can be written in terms of the activation free energy for an equilibrium liquid at a higher specific configurational entropy:
\begin{eqnarray}
\Delta F^\ddag_{n.e.} = \Delta F^\ddag_{eq} (s_c + \frac{\Delta \Phi}{k_B T}) \label{eq:Fne2}
\end{eqnarray}
$\Delta F^\ddag_{eq}(s_c)$ is the function giving the activation barrier in the liquid, written in terms of its configurational entropy. In the glass below its $T_g$  this formula implies the rates follow something close to an Arrhenius law but with an activation enthalpy diminished from what it was at the higher temperature at which it fell out of equilibrium. In this way, this RFOT argument accurately predicts the so-called nonlinearity parameter $x$ describing the ratio of activation enthalpy for motion in an equilibrated liquid to that for glasses that have fallen out of equilibrium \cite{lubchenko2004theory}. 

The Volger-Fulcher law, while describing the deep glassy behavior, breaks down at higher temperatures in supercooled liquids. In mean field theory this breakdown occurs because the mismatch coefficient  $\gamma$ itself vanishes at the mean field spinodal temperature $T_A$ \cite{kirkpatrick1987stable,cammarota2009evidence,lubchenko2003barrier}. Schmalian, Stevenson and Wolynes have argued that the Volger-Fulcher relation will actually break down at a somewhat lower temperature $T_c$ because the shape of correlated activated regions changes at higher temperatures in such a way that the mismatch energy now scales linearly in $N$ \cite{stevenson2006shapes}. The high temperature, entropically favored shapes are lattice animals whose exposed surface scales directly with their number of constituents as does their shape entropy. In this regime the scaling of mismatch energy can be written as  $v_{int} b$   where $b$ is the number of equivalent broken bonds at the surface of the rearranging region. Schmalian, Stevenson and Wolynes showed that near $T_K$, $b$ is approximately $3.2 N$ and the coefficient of  $v_{int}$ can (like the surface mismatch energy $\gamma$) be obtained from density functional reasoning $v_{int}  = (1/z) (3/2) k_B T \ln (\alpha_L a^2/\pi e) $.

The free energy profile for such fractal rearranging regions either monotonically  increases with $N$ or decreases monotonically with $N$. This means that there will be a change in the rearrangement mechanism from activated dynamics to one dominated by collisions at high temperature. The crossover to barrierless reconfiguration is thus determined by the condition 
$s_c (T_c) =  s_c^{\mathrm{perc}} - \Delta \Phi/T_c = \Delta c_p(T_g) T_g/T_K (1-T_K/T_c)\ $. For equilibrated liquids this relation predicts crossover temperatures agreeing with those found using Stickel plots \cite{stickel1996dynamics}. The RFOT argument then also predicts barrierless reconfiguration will occur for nonequilibruim glasses if heated when
\begin{eqnarray}
\frac{k_B T_c}{\Delta c_p(T_g) T_g - \Delta \Phi} = \frac{k_B}{\Delta c_p(T_g)}\frac{T_K}{T_g} \left( 1 - \frac{s_c^{perc}}{\Delta c_p}\frac{T_K}{T_g} \right)^{-1}.
\end{eqnarray}
This specific prediction for crossover to collisional dynamics in superheated glasses has not yet been tested in the laboratory although it would be interesting to check it experimentally by using lasers to rapidly heat glasses.

What does RFOT theory then suggest about how reconfiguration events occur under stress? If a sample of glass is put under a uniform shear stress, $\sigma$, the energy per unit volume is immediately raised by an amount $\sigma^2/2 G $ where $G$ is the elastic shear modulus \cite{sausset2010solids}. It has long been known this energy can be explosively released by cracking the glass. This effect is demonstrated by the famous Prince Rupert's drops {\cite{hooke1665observation}. The stress need not always lead to cracking directly. It is reasonable to expect that, of the myriad of possible states envisioned in RFOT theory, a significant fraction will also allow this stress energy to be released without cracking or forming voids. Indeed a vanishing stress energy of the rearranged state is expected since the most stable mean field free energy minimum corresponding to delocalized molecules can be thought of as being a typical disordered liquid ensemble and is thus  completely  incapable of sustaining static shear. Since the imposed stress energy can be removed by appropriately rearranging a region, imposed strains will lower the activation barrier and will catalyze the rearrangement. If the stress is sufficiently large the rearrangement may even occur without any significant barrier at all, just as takes place at the thermal crossover at $T_c$. This crossover to barrierless reconfiguration would thus give the limiting strength of the glass if we assume there are no easier routes for the glass to rearrange (like cracking). Strain catalysis means that a stressed glass will always flow at some finite rate even for the smallest stresses \cite{argon1979plastic,sausset2010solids} and thus a glass will deform, given time, at somewhat lower stresses than this limit. This gentler flowing situation is probably quite relevant in many practical situations. Flow itself can act to further catalyze rearrangements. The resulting additional enhancement of reconfiguration speed is a facilitation or mode coupling effect. Lubchenko has shown that this effect (that would be contained in a more complete RFOT theory including mode coupling effects \cite{bhattacharyya2008facilitation}) does a good job describing the crossover to steady state nonlinear rheology \cite{lubchenko2009shear}. Similar effects have been studied in mode coupling treatments of dense colloid rheology \cite{fuchs2002theory,brader2009glass}. We will, in this paper, however, concentrate on the immediate effect of stress on the activated events that occur before flow starts and leave the physics of developed plastic flows for future work. The limiting strengths we calculate in this paper then should be upper bounds representing extremely rapid failure of the glass.

\begin{figure}[h!] 
  \centering
      \noindent\includegraphics[width=19pc,angle=0]{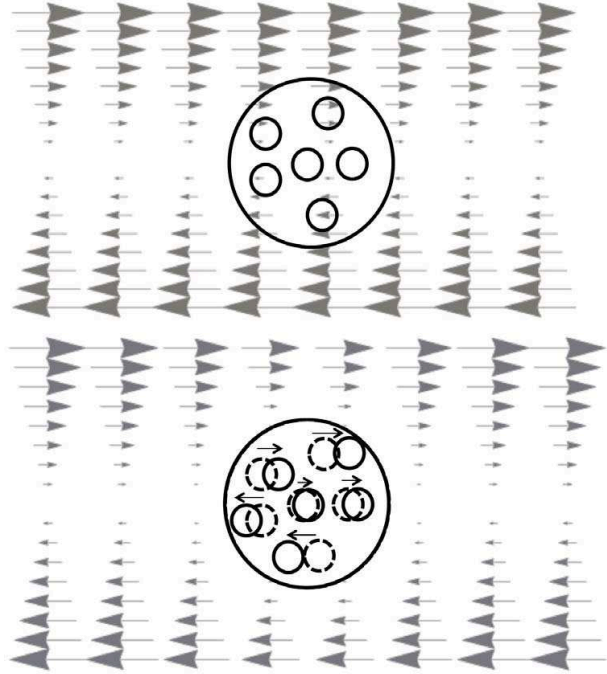} 
  \caption{The upper panel shows a uniform strain field acting on the glass sample in its original state. A fluidized region allows the surrounding material to elastically deform in a nonuniform way in the imposed stress field shown schematically in the lower panel. This allows additional strain energy to be released without costing any additional mismatch energy.} \label{fig:ShearStressField_BW}
\end{figure}

Naively speaking, in order to compute the effect of stress catalysis on the activation barrier one merely needs to account for the strain energy lost in the fluidized region and thus must add to the bulk thermodynamic driving force term $(-T s_c - \Delta \Phi )N$ an additional contribution $\sigma^2N V_{bead}  /2 G$ to compute the lowering of the thermal barrier for compact clusters or to find the limiting stress where barrierless rearrangement may occur. Here $V_{bead}$ is the volume of a ``bead'' i.e. movable unit of the glass, which can be inferred from the molar fusion entropy \cite{lubchenko2003barrier,stevenson2005thermodynamic}. There is a subtlety, however; as pictured in Fig. \ref{fig:ShearStressField_BW}: fluidizing a region of the glass also allows Hookean elastic rearrangements of the surrounding matrix to occur without it being necessary for the outside region to move to any alternate free energy minima. This outer region while elastically responding thus does not elicit any mismatch energy. The strain energy relieved by rearrangement of a region of size $N$ \cite{mansfield2007comparison} nevertheless becomes larger than $\sigma^2 N V_{bead}/2 G$.

For an arbitrarily shaped rearranging region the exact calculation of the additional strain energy relieved by harmonically distorting the outer region would seem to be a  complex problem in elastic theory. The result for spherical regions, however, has been known for some time where it has been used to develop the theory of the elastic modulus of composite media containing holes \cite{mackenzie1950elastic}. The calculation is mathematically quite analogous to the calculation of intrinsic viscosity made first by Einstein \cite{einstein1906calculation} for the effect on viscosity of suspending solid colloidal particles in a fluid and still more closely follows Taylor's analysis of the viscosity changes due to suspending liquid drops or bubbles in a fluid \cite{taylor1923motion}. For spheres the additional energy released (analogous to the excess viscous dissipation in the hydrodynamic problems) is still proportional to the sphere's size and according to MacKenzie depends on the Poisson's ratio characterizing bulk versus pure shear deformations. Taking over MacKenzie's correction gives then an energy increment for rearranging a region of size $N$, $\Delta E_{\mathrm{elastic}} = \kappa \frac{\sigma^2}{2 G} N V_{\mathrm{bead}}$ where $\kappa = 3 - 6/(7 - 5 \nu )$      in terms of the Poisson's ratio $\nu$. For the typical  Poisson's ratio of metallic glasses $\kappa \approx 1.8$. 

Directly calculating the correction for non-spherical shapes is indeed complex. In addition, considerable distortions from spherical geometry are energetically still more favorable in relieving stress than for the region to remain compact and thus at high stress such distortion should again lead to barrierless breakup, just as critical flow rates lead to dissolution of drops in emulsions \cite{taylor1932viscosity,taylor1934formation,stone1994dynamics}. The latter problem has led to an extensive literature \cite{eshelby1959elastic,mansfield2007comparison}. We will use, nevertheless, the spherical value of the correction $\kappa$ for all shapes of cooperatively rearranging regions. We suspect this simplification is  probably not too bad for small stresses and not too far from $T_K$. This surmise is buttressed by the experience for the corresponding hydrodynamic problem of computing the intrinsic viscosity of complex shapes, a problem that has been extensively studied in polymer chemistry \cite{mansfield2007comparison}. In that problem the shape effects are quite modest. By adding the increased relieved strain energy to the reconfiguration driving force, in analogy with equation (\ref{eq:Fne2}), the activation barrier for flow in a strained glass can again be written in terms of the function giving the barrier for equilibrated liquids $\Delta F^\ddag = \Delta F^\ddag \left( s_c + \Delta \Phi/T + \kappa \sigma^2 V_{bead}/2 G T \right)$. With this simplification then barrierless reconfiguration should finally occur when $s_c (T_c) =  s_c^{\mathrm{perc}} - \frac{\Delta \Phi}{T} - \kappa \frac{\sigma^2}{2 G T}V_{bead}$ .  As in the popular J-point scenario \cite{liu1998nonlinear} an apparent  spinodal to reconfiguration apparently can be approached by tuning either the temperature $T$ or the stress $\sigma$.

The argument relating barriers in the glass under stress to those for thermal motions in the equilibrated liquid should hold for temperatures not too far from $T_K$ since the shapes of rearranging regions are determined entropically. There are corrections, however, away from $T_K$. At very high temperatures near to the mean field spinodal  $T_A$ the mismatch energy cost goes down, leading to an additional weakening of the glass. Conversely at low temperatures, much below $T_K$, we also must account for both the fact that the mismatch energy becomes pinned at its $T_K$ value and that at the same time the importance of shape entropy is lessened by the diminished temperature. A detailed account of the latter effects is contained in the Supplementary Information. When the latter effects are included along with the calculation of the excess energy we find an explicit equation for the limiting strength $\sigma^*$:
\begin{eqnarray}
\sigma^*_{\text{pred}} =  \sqrt{ \frac{2 G k_B T}{\kappa V_{bead}} \left( \left[3.20 \frac{T_K}{T} - 1.91 \right]  -  \frac{\Delta c_p(T_g)}{k_B} \frac{T_g}{T}\ln \frac{T_g}{T_K}  \right)}
\end{eqnarray}

The contribution in this expression involving $\Delta c_p $ represents the weakening caused by the excess energy which has been frozen in at the glass transition. If we could be cosmologically patient this excess energy would disappear by annealing to $T_g  = T_K$ giving then the ultimate achievable strength of a glass. At very low temperatures the strength of this most stable glass will be then  $\sigma^*_{ideal} = \sqrt{3.6 G k_B T_K/V_{bead}}$ .  We can write the shear modulus in terms of the spinodal temperature $T_A$ and the bead size, following Lubchenko by estimating vibrational displacements from Debye continuum theory and assuming a limiting Lindemann ratio for the maximal thermal excursions \cite{rabochiy2012liquidy,greaves2011poisson}. Using this relation gives $G = 24.9 k_B T_A/V_{bead}$. If we now take $T_A \approx T_m$ the melting point and use the typical ratio of $T_K/T_m$ of between $0.4$ and $0.7$ we find the ultimate limiting $\sigma^*$ is proportional to the elastic modulus. Such a linear correlation between strength and modulus, resembles Frenkel's estimate, and indeed has been examined experimentally. We find the ideal limit strength from RFOT theory to be uniformly about $30 \%$ higher than Frenkel's. The weakening of the glass due to energy frozen in at the glass transition is however substantial. This excess energy lowers the strength quite a bit below the RFOT ideal value and below the Frenkel value but still gives strengths greatly exceeding the measured strength of polycrystalline metals. We have gathered from the literature data for the input thermodynamics. We then compared the  RFOT theory predictions to measured strengths for some metallic glasses, silica and a polymer glass, PMMA. Details of the input data and measured strength data can be found in the Supplementary Information.

\begin{figure}
  \begin{subfigure}[b]{0.5\textwidth}
  	\centering
	\includegraphics[width=19pc]{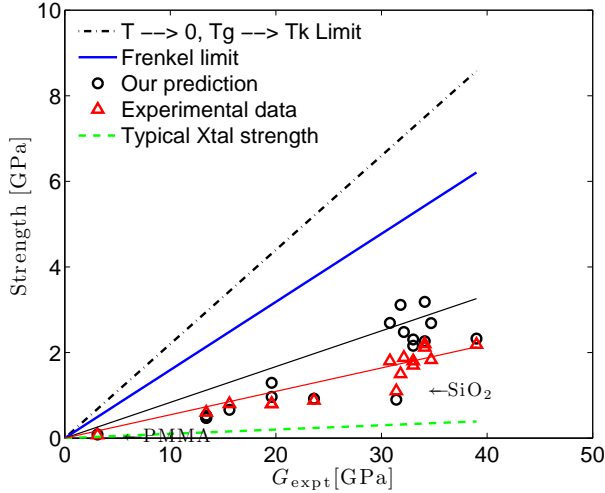} 
	\caption{Strength as a function of shear moduli \\ \ }
	\label{fig:Fig3_SigmaVsModuli}
  \end{subfigure}
   \begin{subfigure}[b]{0.5\textwidth}
   	\centering
	\includegraphics[width=19pc]{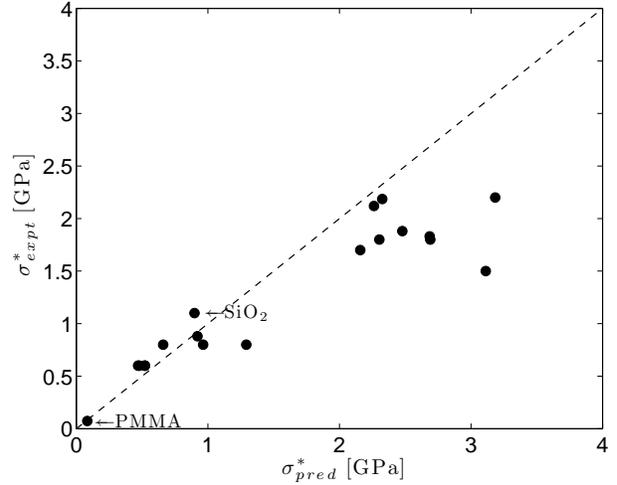} 
	\caption{A plot of the strength $\sigma^*$,measured from the experiments versus theoretical estimation}
	\label{fig:SigmaPredVsExpt2}
  \end{subfigure}
  \caption{The experimental strength (red triangle) and the predicted strength (black circle) have nearly the same slope and are quite different from Frenkel strength (blue solid line). Typical value of crystal strength (green dashed line) and strength in the limit $T \rightarrow 0, T_g \rightarrow T_K$ (black dashed-dotted line) are also shown in comparison.} \label{fig:Strength2}
\end{figure}

In Fig.~\ref{fig:Fig3_SigmaVsModuli} we display results for the strength versus shear modulus. The predicted strengths generally exceed but are close to the measured yield strengths. On this plot we also show both the Frenkel estimate and our $Tg = T_K$ ideal value. A typical polycrystalline material value of one-hundredth of the  Frenkel value is also plotted. Clearly the present RFOT predictions account very well not only for the trends but even the actual magnitude of the strength. In Fig.~\ref{fig:SigmaPredVsExpt2} we show the comparison of measured strengths against the complete predictions. Not only the glassy metals but also silica and the plastic PMMA have strengths not terribly far off the RFOT predictions.   Although the main dependence on elastic modulus is clear, the RFOT theory results also depend on other quantities, such as $\Delta c_p$ and the ratio of the ambient temperature to glass transition temperature. However, as we can see, both the predicted ratios of strength to modulus for the measured systems and the measured ratios show no overall trend with liquid fragility or glass transition temperature, see Fig.~S1 in the Supplementary Information. Of course since modulus and $T_g$ are well correlated the absolute strengths themselves do correlate with $T_g$.  It may be possible to test the theory further. Again, rapid heating should lower the yield strength in a predictable way. In addition, superstable glasses can be made via vapor deposition \cite{Swallen19012007}. Their effective temperature corresponds to being roughly half way to the Kauzmann temperature. We see their strength should thus be proportionately closer to the Frenkel limit.

RFOT theory accounts well for the measured strength of laboratory glasses of various composition. The good agreement between theory and experiment suggests that the correlated rearranging regions responsible for high temperature viscosity in supercooled liquids also limit the strength of nonequilibrium glasses. There seems to be no necessity to invoke then any additional defects of a point-like or line-like character to play a prominent role in weakening glasses that are prepared in an ordinary fashion by cooling a melt.

\section*{Acknowledgement}
P.G.W. wishes to thank Prof. A. L. Greer and Prof. G. N. Greaves for stimulating conversations. He also very much appreciates the hospitality of the Chemistry Department and Sidney-Sussex College of Cambridge University where this work was begun. Financial support by the D.R. Bullard-Welch Chair at Rice University and a Royal Thai Government Scholarship to A.W. are gratefully acknowledged.

\newpage
\bibliographystyle{naturemag}
\bibliography{OnStrengthOfGlasses}


\end{document}


\title{Supplementary Information \\ On the Strength of Glasses}
\author{Apiwat Wisitsorasak$^{a,b}$ and Peter G. Wolynes$^{*,a,b,c}$ \\
$^a$ \small Center for Theoretical Biological Physics, $^b$ \small Department of Physics \& Astronomy, \\
$^c$ \small Department of Chemistry, Rice University, Houston, TX \\
\texttt{pwolynes@rice.edu}}
\maketitle

\renewcommand{\theequation}{S\arabic{equation}}
\renewcommand{\thefigure}{S\arabic{figure}}
\renewcommand{\thetable}{S\arabic{table}}
\section{Derivation of equation~(4)}
\setcounter{page}{14}

The free energy cost for cooperatively rearranging a region (CRR) of $N$ sites with $b$ boundary interaction is 
\begin{eqnarray}
F(N,b,\sigma) = [f_{eq}(T) - \phi_{in}(T,\sigma)] N + v_{int}b - k_B T \ln\Omega(N,b) \label{eq:FreeEnergy1}
\end{eqnarray}
where $\Omega(N,b)$ is the number of lattice animals of given $N$ and $b$, and $s_c$ is the configurational entropy per site. 
The initial nonequilibrated free energy has an additional elastic energy due to stress
\begin{eqnarray}
\phi_{in}(T,\sigma) = \phi_{in}^0(T) + \kappa \frac{\sigma^2}{2 G} V_{bead}
\end{eqnarray}
where $\phi_{in}^0$ is the sum of the energetic and vibrational  entropic contribution to the initial nonequilibrated state.\\
The equilibrated free energy at temperature $T$ above $T_K$ is
\begin{eqnarray}
f_{eq}(T) & = & \phi_K - \int_{T_K}^T dT' s_c(T')  \\
& = & \phi_K - \Delta c_p(T_g) T_g \left( \frac{T-T_K}{T_K} - \ln{\frac{T}{T_K}} \right)
\end{eqnarray}
where we have used Angell's empirical form: $\Delta c_p(T) = \Delta c_p(T_g)(T_g/T)$, and $\displaystyle s_c(T) = \int_{T_K}^T dT' \frac{\Delta c_p(T')}{T'} = \Delta c_p(T_g) T_g \left( \frac{1}{T_K} - \frac{1}{T} \right)$. Note that $\displaystyle f_{eq}(T_g) = \phi_g - T_g s_c(T_g)$, the ideal glass state energy is equal to 
\begin{eqnarray}
\phi_K = \phi_g - \Delta c_p(T_g) T_g \ln \frac{T_g}{T_K}. \label{eq:phiKphig}
\end{eqnarray}
Consider the first term on the right-hand side of equation~(\ref{eq:FreeEnergy1}) and let $\phi_{in}^0(T)$ be the bulk energy at $T_g$. Then use the relation in equation~(\ref{eq:phiKphig}) to obtain
\begin{eqnarray}
f_{eq}(T) - \phi_{in}(T,\sigma) & = & - \Delta c_p(T_g) T_g \left\{  \frac{T-T_K}{T_K} + \ln{\frac{T_g}{T}} \right\} - \kappa  \frac{\sigma^2}{2 G} V_{bead}
\end{eqnarray}
Substitute this result back into equation~(\ref{eq:FreeEnergy1})
\begin{eqnarray}
F(N,b,\sigma) = \left[ - \Delta c_p(T_g) T_g \left\{  \frac{T-T_K}{T_K} + \ln{\frac{T_g}{T}} \right\} - \kappa  \frac{\sigma^2}{2 G} V_{bead} \right] N + v_{int}b - k_B T \ln\Omega(N,b) .
 \end{eqnarray}
 If $T$ is below $T_K$, the excess energy is frozen at the state $T = T_K$ and the configurational entropy vanishes.
\begin{eqnarray}
F(N,b,\sigma) = \left[ - \Delta c_p(T_g) T_g  \ln{\frac{T_g}{T_K}} - \kappa  \frac{\sigma^2}{2 G} V_{bead} \right] N + v_{int}b - k_B T \ln\Omega(N,b).\label{eq:FreeEnergy2}
 \end{eqnarray} 
 
 In percolation clusters \cite{leath1976cluster}, for large $N$, 
 \begin{eqnarray}
 \Omega_{\mathrm{perc}} \equiv \left(\frac{(\alpha+1)^{\alpha+1}}{\alpha^\alpha} \right)^N \exp \left(- \frac{N^{2 \phi}}{2 B^2} (\alpha - \alpha_e)^2 \right)
 \end{eqnarray}
 where $\alpha = t/N$, and $t$ is the number of unoccupied sites bounding the occupied cluster. Take the exponent $\phi$ at mean field value of $1/2$ and a lattice constant $B = 1.124$.
 The number of bonds $b$ is directly related to $t$ and should linearly depend on coordination number, $z$, $b \approx 1.68 t z/z_{\mathrm{SC}} \cite{stevenson2005thermodynamic}$. Recall that the surface energy $\displaystyle v_{int} = \frac{1}{z} T_K \left(\frac{3}{2} k_B \ln \left(\frac{\alpha_L a^2}{\pi e} \right) \right) = 3.6907 \frac{k_B T_K}{z}$. The free energy in equation~(\ref{eq:FreeEnergy2}) is now a function of $N$ and $t$. Minimize this function with respect to $t$, the most probable value of $t$ is $\bar t = \bar \alpha N$, where $\bar \alpha = 3.10$ at $T=T_K$. At this most probable value, $\Omega_{\mathrm{perc}}$ becomes a simple exponential function, $\Omega_{\mathrm{perc}} \sim \lambda^N$, where $\lambda = 6.82$. Each term in  equation~(\ref{eq:FreeEnergy2}) is now proportional to $N$ and we can write
 \begin{eqnarray}
F(N,b,\sigma) &=& \left[ - \Delta c_p(T_g) T_g  \ln{\frac{T_g}{T_K}} - \kappa  \frac{\sigma^2}{2 G} V_{bead} \right] N \nonumber \\
 & & + v_{int} 1.68 \frac{z_{fcc}}{z_{SC}} \bar \alpha N   - k_B T N \ln \lambda  \\
 &=&  k_B T N\left\{ \left[ - \frac{\Delta c_p(T_g)}{k_B} \frac{T_g}{T} \left\{\ln{\frac{T_g}{T_K}} \right\} 
 - \frac{1}{k_B T}\frac{\kappa \sigma^2}{2 G} V_{bead} \right] \right. \nonumber  \\
 & &  \left. + \frac{v_{int}}{k_B T} 1.68 \frac{z_{f.c.c.}}{z_{SC}} \bar \alpha - \ln \lambda \right\} 
 \end{eqnarray}
 At the thresholding stress $\sigma^*$ where $F(N,\sigma^*) = 0$, one finds
 \begin{eqnarray}
 \sigma^* = \sqrt{ \frac{2 G k_B T}{\kappa V_{bead}} \left( \left[\frac{v_{int}}{k_B T} 1.68 \frac{z_{f.c.c.}}{z_{SC}} \bar \alpha - \ln \lambda \right]  -  \frac{\Delta c_p(T_g)}{k_B} \frac{T_g}{T} \ln{\frac{T_g}{T_K}} \right)}, \label{eqn:SigmaPred1}
\end{eqnarray}
or
 \begin{eqnarray}
 \sigma^* = \sqrt{ \frac{2 G k_B T}{\kappa V_{bead}} \frac{v_{int}}{k_B T} \left( 1.68 \frac{z_{f.c.c.}}{z_{SC}} \bar \alpha - \left[ \ln \lambda +   \frac{\Delta c_p(T_g)}{k_B} \frac{T_g}{T} \ln{\frac{T_g}{T_K}} \right]\frac{k_B T}{v_{int}}    \right)}.
\end{eqnarray}
Substituting the numbers in the previous paragraph, we finally obtain equation~(4)
\begin{eqnarray}
\sigma^* = \sqrt{ \frac{2 G k_B T}{\kappa V_{bead}} \left( \left[3.20 \frac{T_K}{T} - 1.91 \right]  -  \frac{\Delta c_p(T_g)}{k_B} \frac{T_g}{T} \ln{\frac{T_g}{T_K}} \right)}.
\end{eqnarray}

\section{Experimental data and numerical prediction}
In Table S1 we summarize the input thermodynamic data and measured strengths as well as their sources in the literature. The bead count is obtained as described by Lubchenko and Wolynes \cite{lubchenko2003barrier} and Stevenson and Wolynes \cite{stevenson2005thermodynamic} using the melting entropy. All strength measurements were all made at room temperature $300~ K$.
\begin{landscape}
\begin{table}[position specifier]
\centering
\begin{tabular}{ c c c c c c c c c c c c c} \hline
 Glasses & $\rho$  & $\Delta H_M$ & $N_{\text{bead}}$ & $T_K$ & $T_g$ & $T_M$ & $\Delta c_p$ & $G_{\mathrm{expt}}$ & $\sigma^*_{\text{expt}}$& $\sigma^*_{\text{pred}}$& Refs. \\ \hline
  PMMA & 1188 &  4.64 & 0.84 & 337 & 372 & 397 & 6.68 & 3.10 & 0.07 & 0.08 & \cite{ute1995glass,fernandez1981glass,prevosto2003correlation,liberman1984compatibility,lee2009direct,jang1999mechanical,guo2012ultrastable} \\
  \cf{SiO2} & 2648  & 9.6 & 0.78 & 876 & 1500 & 1995 & 1.67 & 31.40 & 1.1 & 0.90 & \cite{richet1984viscosity,rabochiy2012liquidy,bridge1983elastic,northolt1981compressive,bacon1977high} \\
  \cf{Mg65Cu25Y10} & 3978  & 8.65 & 0.66 & 325 & 424 & 771 & 2.41 & 19.60 & 0.8 & 1.29 & \cite{battezzati2007glass,busch:4134,cai2007relationship,wang2011elastic,sheng2005evaluation,cai2008estimation,yang2006localized,jiang2007intrinsic} \\  
  \cf{Mg65Cu20Zn5Y10} & 3284  & 7.77 & 0.79 & 325 & 404 & 702 & 2.38 & 23.60 & 0.88 & 0.92 & \cite{cai2007relationship,hui2006situ} \\    
  \cf{Mg80Cu10Y10} & -  & 7.21 & 0.69 & - & 427 & 746 & 2.73 & - & 0.8 & 0.96 & \cite{cai2007relationship,murty2000formation,inoue2001high} \\
  \cf{La55Al25Ni20} & 6140  & 7.48 & 0.75 & 337 & 491 & 712 & 2.15 & 13.40 & 0.6 & 0.52 & \cite{cai2008estimation,lu2003glass,perera1999compilation,cai2007relationship,yang2006localized} \\
  \cf{La55Al25Cu5Ni15} & 6050  & 7.51 & 0.82 & 328 & 472 & 660 & 2.21 & - & 0.6 & 0.47 & \cite{lu2003glass,cai2007relationship,cai2008estimation} \\
  \cf{La55Al25Cu10Ni10} & 5930  & 6.84 & 0.74 & 332 & 467 & 662 & 2.40 & - & 0.6 & 0.48 & \cite{lu2003glass,cai2007relationship,cai2008estimation} \\
  \cf{La55Al25Cu15Ni5} & 6370  & 7.21 & 0.78 & 320 & 459 & 663 & 2.14 & - & 0.6 & 0.52 & \cite{lu2003glass,cai2007relationship,cai2008estimation} \\
  \cf{La55Al25Co5Cu10Ni5} & 6000  & 6.09 & 0.66 & 363 & 466 & 661 & 2.84 & 15.60 & 0.8 & 0.67 & \cite{lu2003glass,cai2007relationship,cai2008estimation,johnson2005universal} \\
  \cf{Pd40Ni40P20} & 8951  & 7.39 & 0.60 & 487 & 570 & 884 & 3.28 & 39 & 2.19 & 2.33 & \cite{battezzati2007glass,hu1999glass,cai2007relationship,yu2010electronic,cai2008estimation,yang2006localized} \\
  \cf{P40Cu30Ni10P20} & 9300  & 6.84 & 0.61 & 497 & 593 & 798 & 3.39 & 33.00 & 1.8 &   2.30 & \cite{battezzati2007glass,lu2000correlation,cai2007relationship,cai2008estimation,johnson2005universal,yang2006localized} \\
  \cf{Pd77Cu6Si17} & 10400  & 8.55 & 0.58 & 553 & 637 & 1058 & 2.74 & 31.80 & 1.5 & 3.11 & \cite{komatsu1995application,cai2006evaluation,wang2011elastic,lu2002new} \\
  \cf{Zr11Cu27Ni8Ti34} & 6850  & 11.3 & 0.72& 537 & 671 & 1128  & 2.25 & - & 2.2 &  3.20 & \cite{battezzati2007glass,bossuyt2001microstructure,glade:7242,cai2008estimation,lu2002new} \\
  \cf{Zr_{41.2}Ti_{13.8}Ni10Cu_{12.5}Be_{22.5}} & 6125  & 8.2 & 0.63 & 553 & 620 & 937 & 4.03 & 34.10 & 2.12 & 2.26 & \cite{battezzati2007glass,busch1995thermodynamics,cai2007relationship,wang2004bulk,cai2008estimation,wang2000crystallization,johnson2005universal} \\
  \cf{Zr_{46.25}Ti_{8.25}Ni10Cu_{7.5}Be_{27.5}} & 6014  & 9.4 & 0.57 & 550 & 622 & 1185 & 3.60 & 34.70 & 1.83 & 2.68 & \cite{battezzati2007glass,busch:4134,cai2006evaluation,wang2004bulk,wang2011elastic,demetriou2009rheology,wiest2010thermoplastic,lu2002new} \\
  \cf{Zr_{52.5}Cu_{17.9}Ni_{14.6}Al10Ti5} & 6730  & 8.2 & 0.55 & 633 & 675 & 1072 & 4.35 & 32.12 & 1.88 & 2.48 & \cite{cai2008estimation,glade:7242,cai2007relationship,wang2011elastic,wang2000crystallization,pang2012valence} \\
  \cf{Zr57Cu_{15.4}Ni_{12.6}Al10Nb5} & 6690  & 9.4 & 0.62 & 656 & 682 & 1091 & 3.26 & 30.80 & 1.8 &  2.69 &  \cite{cai2008estimation,glade:7242,cai2007relationship,wang2011elastic,pang2012valence} \\
  \cf{Zr65Al_{7.5}Cu_{27.5}} & 6744  & 12.8 & 0.80 & 517 & 666 & 1150 & 2.24 & 33.00 & 1.7 & 2.16 & \cite{cai2008estimation,perera1999compilation,cai2007relationship} \\ \hline
\end{tabular}
\caption{Experimental data and theoretical results of 19 glasses: density $\rho$ ($\mathrm{kg/m^3}$),  latent heat of fusion $\Delta H_M$ ($\mathrm{kJ/mol\ K}$), number of bead $N_{bead}$, Kauzmann temperature $T_K$ ($\mathrm{K}$), glass transition temperature $T_g$ ($\mathrm{K}$), melting temperature $T_M$ ($\mathrm{K}$), heat capacity change at glass transition temperature $\Delta c_p$ ($\mathrm{J/mol\ K}$), experimental shear modulus $G_{expt}$ ($\mathrm{GPa}$), measured strength $\sigma_{expt}^*$ ($\mathrm{GPa}$), and theoretical estimated strength $\sigma_{pred}^*$ ($\mathrm{GPa}$).}
\label{tab:myfirsttable}
\end{table}
\end{landscape}

\begin{figure}[h]
  \centering
      \noindent\includegraphics[width=19pc,angle=0]{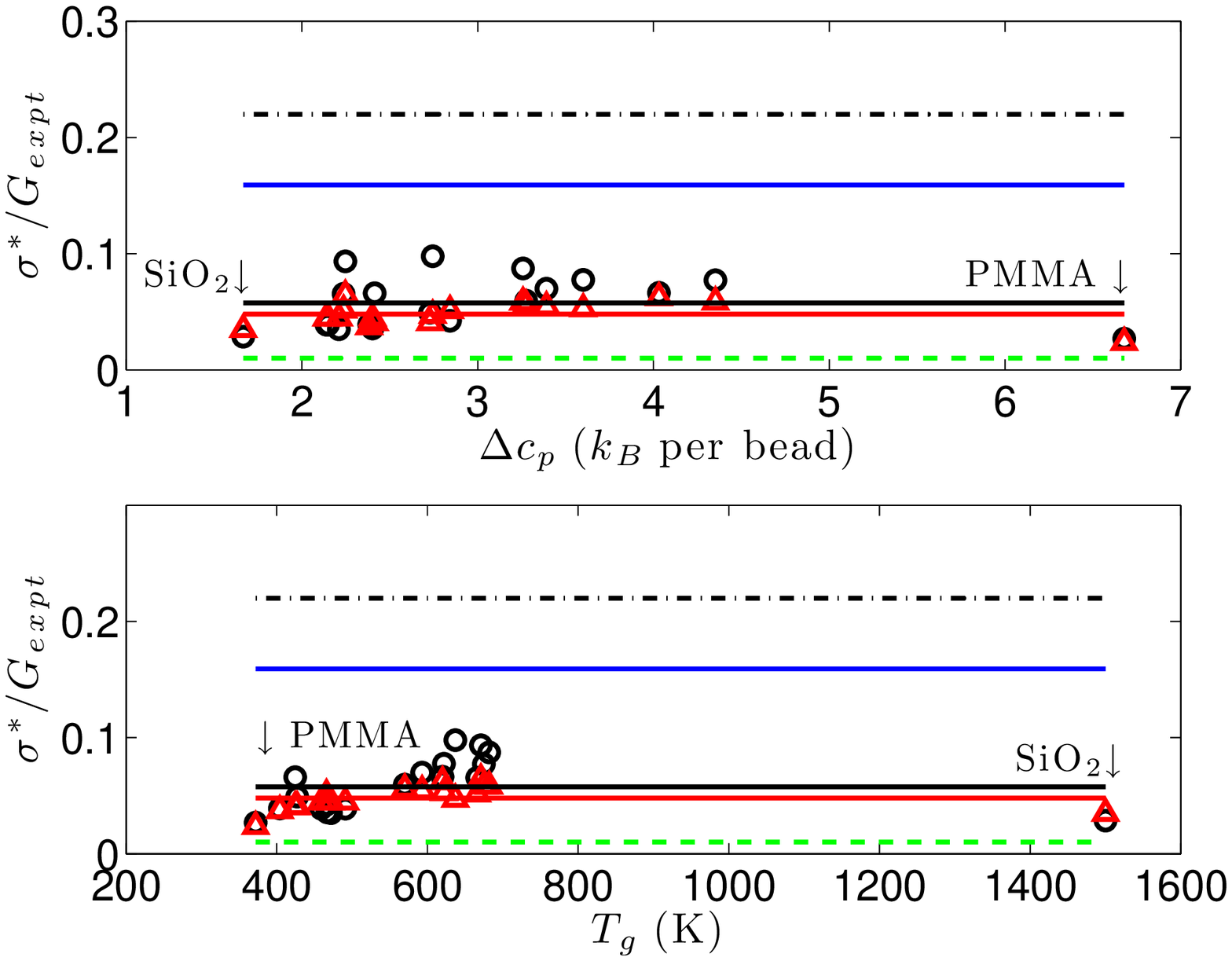} 
  \caption{The ratio between strength and shear moduli versus heat capacity change at $T_g$ and the glass transition temperature $T_g$. The black circles are the RFOT theory predictions and the red circles are the measured values.} \label{fig:SlopePlot2_Tg_Cp}
\end{figure}

\begin{figure}[h]
  \centering
      \noindent\includegraphics[width=19pc,angle=0]{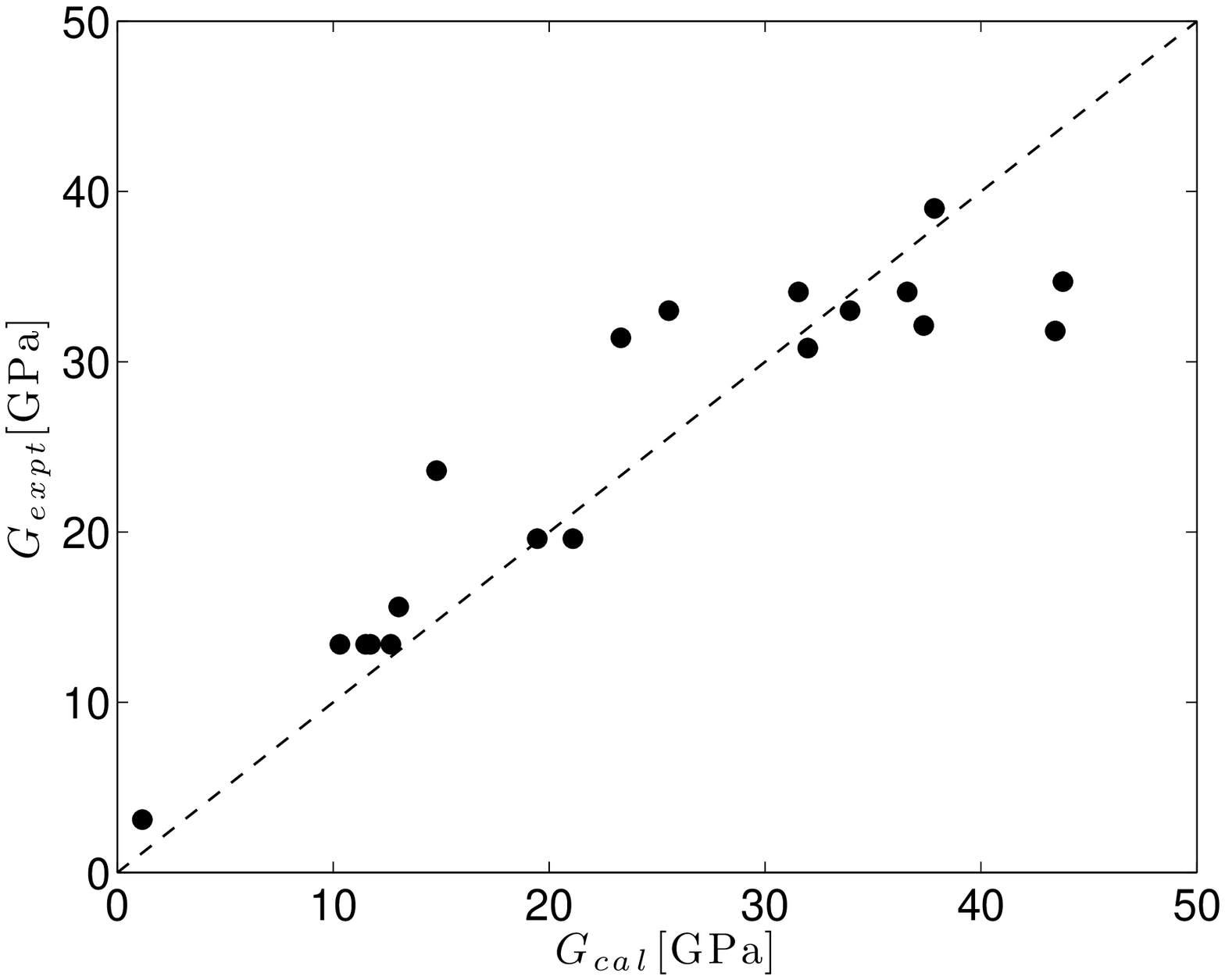} 
  \caption{Comparison between the measured elastic moduli and those calculated using the Lindemann criterion.} \label{fig:Fig_ShearModulus}
\end{figure}

We also show here in Fig.~\ref{fig:Fig_ShearModulus} the comparison of measured elastic moduli with those predicted via the relation $G_{cal} = 24.9 k_B T_m/V_{\mathrm{bead}}$ from thermodynamic data along with the Lindemann relation and the assignment $T_A \approx T_m$. 

\newpage

\bibliographystyle{naturemag}
\bibliography{OnStrengthOfGlasses}